\documentclass[twocolumn,preprintnumbers,amsmath,amssymb]{revtex4}
\usepackage{graphicx}% Include figure files
\usepackage{dcolumn}% Align table columns on decimal point
\usepackage{bm}% bold math
\usepackage{color}
\usepackage{ulem}
\usepackage{setspace}
\begin{document}
\title{Fabrication and magnetic control of Y$_{3}$Fe$_{5}$O$_{12}$ cantilevers}

\author{Yong-Jun Seo$^{\, 1,2}$}
\email{seo@imr.tohoku.ac.jp}
\author{Kazuya Harii$^{\, 1,3}$}
\author{Ryo Takahashi$^{\, 1,3}$}
\author{Hiroyuki Chudo$^{\, 1,3}$}
\author{Koichi Oyanagi$^{\, 4}$}
\author{Zhiyong Qiu$^{\, 2}$}
\author{Takahito Ono$^{\, 5}$}
\author{Yuki Shiomi$^{\, 1,4}$}
\author{Eiji Saitoh$^{\, 1,2,3,4}$}

\affiliation{$^{1}$
Spin Quantum Rectification Project, ERATO, Japan Science and Technology Agency, Aoba-ku, Sendai 980-8577, Japan}
\affiliation{$^{2}$
WPI Advanced Institute for Materials Research, Tohoku University, Sendai 980-8577, Japan}
\affiliation{$^{3}$
Advanced Science Research Center, Japan Atomic Energy Agency, Tokai 319-1195, Japan}
\affiliation{$^{4}$
Institute for Materials Research, Tohoku University, Sendai 980-8577, Japan }
\affiliation{$^{5}$
Graduate School of Engineering, Tohoku University, Sendai 980-8579, Japan}

%\date{\today}

%%%%%%%%%%%%%%%%%%%%%%%%%%%%%%%%%%%%%%%%%%%
%%%%%%%%%%%%%%%%%%%%%%%%%%%%%%%%%%%%%%%%%%%
%%%%%%%%%%%%%%%%%%%%%%%%%%%%%%%%%%%%%%%%%%%
\begin{abstract}
We have fabricated ferrite cantilevers in which their vibrational properties can be controlled by external magnetic fields. Submicron-scale cantilever structures were made from Y$_{3}$Fe$_{5}$O$_{12}$ (YIG) films by physical etching combined with use of a focused ion beam milling technique. We found that the cantilevers exhibit two resonance modes which correspond to horizontal and vertical vibrations. Under external magnetic fields, the resonance frequency of the horizontal mode increases, while that of the vertical mode decreases, quantitatively consistent with our numerical simulation for magnetic forces. The changes in resonance frequencies with magnetic fields reach a few percent, showing that efficient magnetic control of resonance frequencies was achieved.
\end{abstract}

%%%%%%%%%%%%%%%%%%%%%%%%%%%%%%%%%%%%%%%%%%%
%%%%%%%%%%%%%%%%%%%%%%%%%%%%%%%%%%%%%%%%%%%
%%%%%%%%%%%%%%%%%%%%%%%%%%%%%%%%%%%%%%%%%%%
\maketitle
%\section{Introduction}
Spin mechanics \cite{spin-mechanics}, which explores interplay between magnetism and mechanical motion, is a young research field emerging along with the advance in spintronics \cite{spintronics}. Classical examples of such phenomena are the Einstein-de Haas effect \cite{EdH} and its inverse effect, the Barnett effect \cite{Barnett}. In the Einstein-de Haas effect, mechanical rotation is induced by transfer of angular momentum from magnetization to mechanical ones. To detect mechanical effects induced by spins, a cantilever structure provides one of the most suitable tools \cite{wallis, mohanty, zolf, boales, Losby, Wu}. A cantilever is a long rigid plate of which one end is supported tightly but the other end can mount a load. Because of their high sensitivity \cite{stowe, chaste}, cheapness, and ease of fabrication in large areas, cantilever structures have been essential in spin mechanics \cite{spin-mechanics, wallis, mohanty, zolf, boales, Losby, Wu}.
\par

In commercial devices {\it e.g.} micro-electro-mechanical systems (MEMS), cantilevers are mostly fabricated on silicon wafers. Silicon is the most common semiconducting material on the earth, and widely used as a base material in the semiconductor industry. Silicon cantilevers thereby have great advantage since nanofabrication techniques developed in the semiconductor industry can be harnessed effectively. However, recent development in state-of-the-art nanofabrication techniques such as a focused ion beam (FIB) method enables wide material choice as ingredients of cantilevers, such as magnetic, piezoelectric, and ferroelectric materials. Cantilevers made of such functional materials are promising for exploration of new features in minute cantilever devices.
\par

In this study, we have fabricated ferrimagnetic cantilevers using garnet ferrite Y$_{3}$Fe$_{5}$O$_{12}$ (YIG). YIG is a typical magnetic insulator \cite{zhang, wago, midzor, charbois, naletov, klein} with excellent microwave properties, and thus has widely been used in magnonics and spintronics fields \cite{spintronics}. However, direct fabrication of YIG cantilevers has not been reported yet, although magnetic control of cantilever properties is expected owing to the strong spontaneous magnetization of YIG. In addition to functionality as magnetic cantilevers, a marriage between MEMS technology and spintronics will acceralate the study of spin mechanics. As shown in the following, we successfully fabricated a YIG cantilever with a Pt mirror {\it in situ} using an FIB milling technique, and demonstrated efficient control of resonant frequencies by using small external magnetic fields. 
\par

\begin{figure}[t]
\begin{center}
\includegraphics[width=8.5cm]{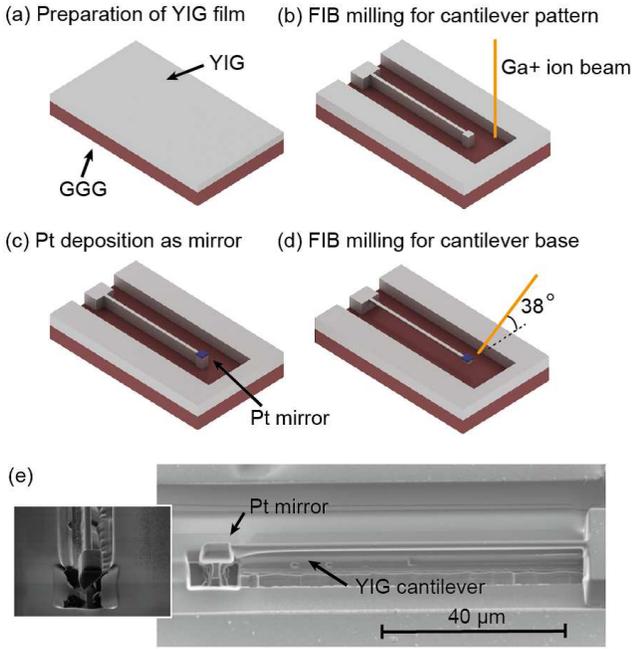}
\caption{(a)-(d) Schematic illustrations of the fabrication process of YIG cantilever. (e) SEM images of the fabricated YIG cantilever. Cross sectional image at the head part is also shown in (e).}
\label{fig1}
\end{center}
\end{figure}

%%%%%%%%%%%%%%%%%%%%%%%%%%%%%%%%%%%%%%%%%%%%
%%%%%%%%%%%%%%%%%%%%%%%%%%%%%%%%%%%%%%%%%%%%
%%%%%%%%%%%%%%%%%%%%%%%%%%%%%%%%%%%%%%%%%%%%
%\section{Experimental details}
Figure \ref{fig1} shows the fabrication process of our YIG cantilever. A YIG cantilever with a Pt mirror was fabricated using a dual beam FIB/SEM system (Versa3D DualBeam; FEI Company). The starting material is a YIG film with $3$ ${\rm \mu m}$ thickness grown on a gadolinium gallium garnet (GGG) substrate. A cantilever structure was patterned by the FIB milling, as shown in Fig. \ref{fig1}(b). The depth of the milling was about $6$ ${\rm \mu m}$, which is much greater than the thickness of the YIG layer. In order to improve the reflectivity of the laser light used in the Doppler vibrometry, a Pt film was deposited on the head of the cantilever {\it in situ} using the FIB deposition, as shown in Fig. \ref{fig1}(c). Then, the base of the cantilever was milled away by the FIB milling at the angle of $38$ degrees from the film plane, as shown in Fig. \ref{fig1}(d). This process was repeated for the other side (Fig. \ref{fig1}(d)), and then the YIG cantilever structure was obtained. A fabricated YIG cantilever with a Pt mirror is shown in Fig. \ref{fig1}(e); the size is $0.8$ ${\rm \mu m}$ in width, $0.9$ ${\rm \mu m}$ in thickness, and $80$ ${\rm \mu m}$ in length. The cantilever is not completely symmetric, as shown in a cross-sectional image in Fig. \ref{fig1}(e). For comparison, non-magnetic Gd$_{3}$Ga$_{5}$O$_{12}$ cantilevers were fabricated using the same method.
\par

\begin{figure}[t]
\begin{center}
\includegraphics[width=8.5cm]{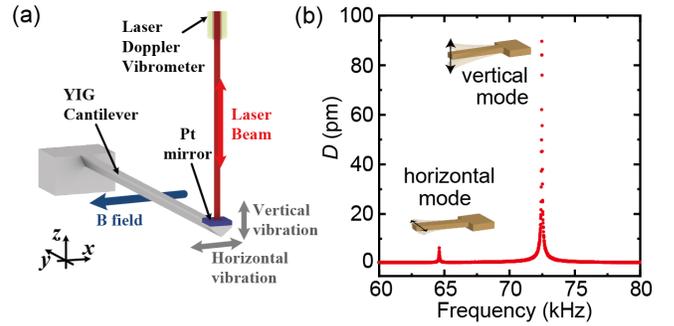}
\caption{(a) A schematic illustration of measurement setup. (b)Frequency dependence of displacement measured along the direction normal to the film plane (denoted by $D$) for the YIG cantilever. In this frequency range, two resonance modes, horizontal and vertical modes, are observed as sharp peaks.}
\label{fig2}
\end{center}
\end{figure}

Vibration spectra of the fabricated cantilevers in the direction normal to the cantilever ($z$ axis in Fig. \ref{fig2}(a)) were measured with a laser Doppler vibrometer (MSA-100-3D; Polytec Inc.) at room temperature, as illustrated in Fig. 2(a). Here, the cantilever vibration is mainly driven by thermal energy of the cantilever, but other minor mechanisms such as residual vibrational/acoustic excitation and electrical noise may exist. The measurement was performed in a high vacuum of $10^{-4}$ Pa to improve sensitivity. External magnetic fields were applied to the cantilever samples using electromagnets along the perpendicular direction to the cantilever within the film plane ($x$ axis in Fig. \ref{fig2}(a)). 
\par

%%%%%%%%  Figure 1  %%%%%%%%%%%

%%%%%%%%%%%%%%%%%%%%%%%%%%%%%%%%%%%%%%%%%%%%
%\section{Results and discussion}
Figure \ref{fig2}(b) shows the frequency dependence of displacement, $D$, of the YIG cantilever. In the frequency range from $60$ to $80$ kHz, two sharp peaks are observed; the frequencies of the peaks are $64.656$ kHz and $72.516$ kHz. From a numerical simulation using COMSOL Multiphysics software \cite{comsol}, we assigned these peaks as resonance modes of the cantilever. The lower resonance frequency ($64.656$ kHz) corresponds to the horizontal vibrational mode, while the higher one ($72.516$ kHz) the vertical vibrational mode. Though the laser beam was set to be perpendicular to the film plane [Fig. \ref{fig2}(a)], small distortion in cantilever shape enables the detection of the horizontal vibration mode. In commercial cantilevers, since the cantilever thickness is much less than the cantilever width, the resonance frequencies of the two modes are significantly different; it is noted that the resonance frequency in cantilevers is known to be proportional to the cantilever thickness. In contrast, since the width and thickness of our YIG cantilever are similar, both the horizontal and vertical modes were observed in the similar frequency range. This argument is supported by the fact that the ratio of the two resonance frequencies ($=72.516$ kHz$/64.656$ kHz) almost coincides with that of the cantilever thickness to the width ($=900$ nm$/800$ nm). 
\par

%%%   Figure 2   %%%

From the displacement measurement without applying magnetic fields shown in Fig. \ref{fig2}, the quality factor ($Q$) of the YIG cantilever was estimated to be $1000$. Using a relation of the minimum detectable force \cite{mforce, mforce2}
\begin{equation}
\delta F_{min}= \sqrt{\frac{4k k_{B}T}{2\pi f_{0} Q}},
\end{equation}
the minimum detectable force from the cantilever size is estimated to be $5 \times 10^{-16}$ N for the horizontal and the vertical resonance mode. Here, $k$, $k_{B}$, $T$, and $f_{0}$ are the spring constant, the Boltzmann constant, the cantilever temperature, and the resonance frequency, respectively. The spring constant $k$ of the YIG cantilever is determined using the relation of $k = EI/L^{3}$, where $E$ is The Young's modulus of YIG \cite{elastic1, elastic2}, $I$ is momentum of inertia simulated from the cantilever dimension using the COMSOL software \cite{comsol}, and $L$ is length of the cantilever. The value of $k$ is calculated to be 6 mN/m. The obtained minimum detectable force shows that highly sensitive YIG cantilevers which can detect forces as small as $100$ aN ($=10^{-16}$ N) were fabricated. This minimum detectable force is much less than that used commercially in the atomic force microscopy (AFM).
\par
 
\begin{figure}[t]
\begin{center}
\includegraphics[width=8.5cm]{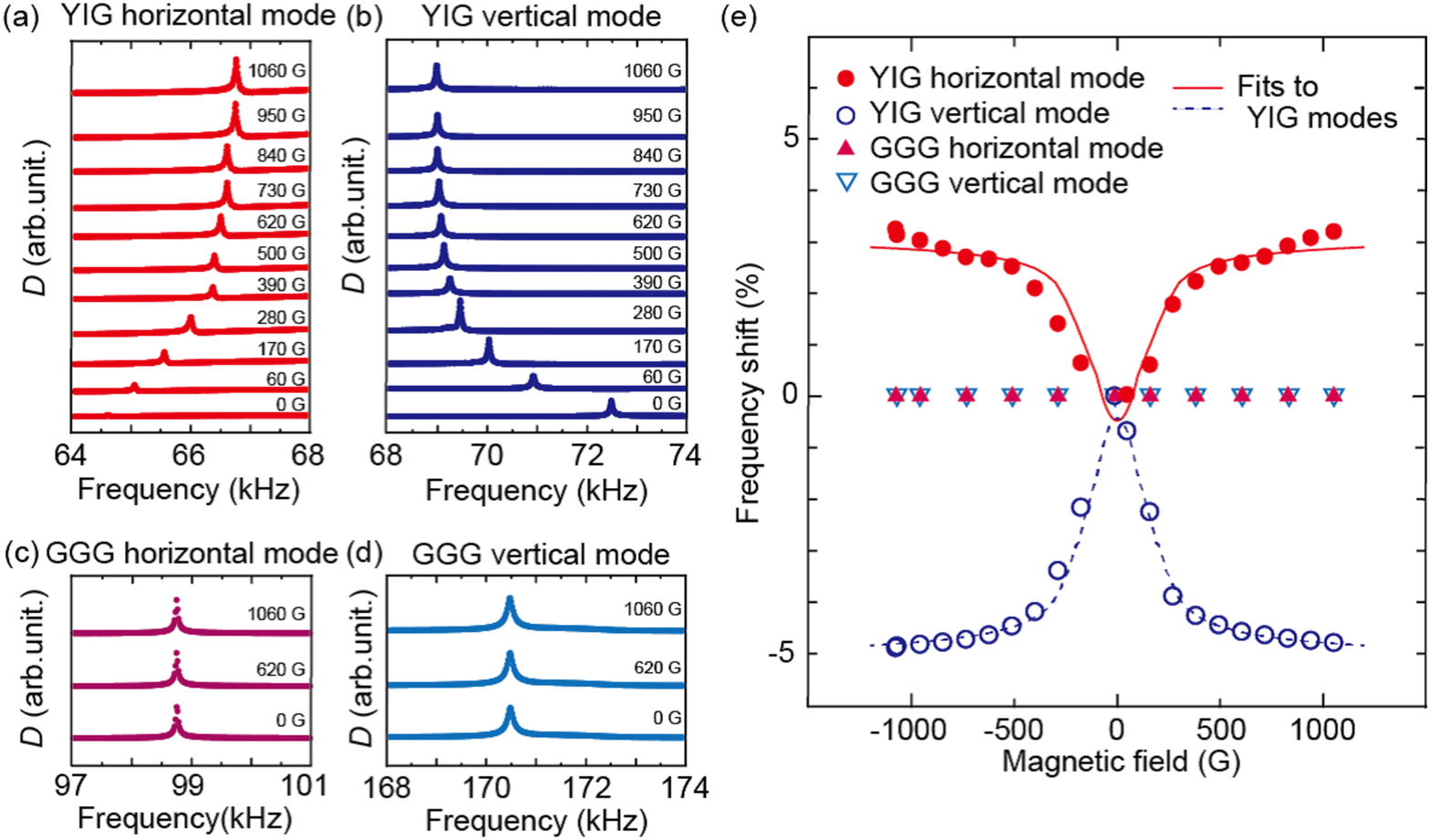}
\caption{Frequency dependence of the displacement (denoted by $D$) measured in the frequency range around the resonance frequencies of (a) horizontal and (b) vertical modes in the YIG cantilever, and (c) horizontal and (d) vertical modes in a non-magnetic GGG cantilever. (e) The frequency shifts with magnetic fields are plotted as a function of external magnetic fields for the YIG and GGG cantilevers. Fits to the experimental results for the YIG cantilever (solid and dashed curves) are also shown.}
\label{fig3}
\end{center}
\end{figure}

Magnetic field dependence of the resonance frequencies is shown in Fig. \ref{fig3}. With increasing magnetic-field strength, the resonance frequencies of both the horizontal and vertical modes change clearly in Figs. \ref{fig3}(a) and \ref{fig3}(b), although the peak shape (i.e. the $Q$ factor) is almost constant with magnetic fields. As shown in Fig. \ref{fig3}(a), the resonance frequency of the horizontal mode steeply increases, as the magnetic field is raised from $0$ G to $390$ G. Above $390$ G, the increase in the resonance frequency with magnetic fields tends to be almost saturated. At $1060$ G, the resonance frequency is $66.753$ kHz, higher than that in zero magnetic field by about $2$ kHz. Also for the vertical mode, the similar strong magnetic field dependence is observed especially at low magnetic fields, as shown in Fig. \ref{fig3}(b). However, on the contrary to the horizontal mode, the resonance frequency of the vertical mode decreases with increasing magnetic fields. Hence, the clear difference in the response to magnetic fields is observed between the two modes.
\par

The magnetic field dependence of the frequency shifts is summarized in Fig. \ref{fig3}(e). The shifts are observed also in negative magnetic fields for the YIG cantilever, and the magnetic field dependence is clearly even with respect to magnetic fields for both the horizontal and vertical modes. The maximal magnitudes of the frequency shifts reach a few percent, indicating that the efficient magnetic control of the resonance frequencies is achieved by small magnetic fields.
\par

%%%  Figure 3  %%%

To examine the origin of the frequency shifts, we performed similar experiments for a non-magnetic Gd$_{3}$Ga$_{5}$O$_{12}$ cantilever. The frequency dependence of $D$ for the GGG cantilever is shown in Figs. \ref{fig3}(c) and \ref{fig3}(d). Also for the GGG cantilever, the resonance modes corresponding to horizontal and vertical modes are observed; the resonance frequencies are different from those for the YIG cantilever because the cantilever sizes are a little bit different. As shown in Figs. \ref{fig3}(c) and \ref{fig3}(d), under external magnetic fields up to $1000$ G,  no frequency shifts are observed either for the horizontal or vertical mode in the GGG cantilever. This results show that the frequency shifts observed in the YIG cantilever are related to the spontaneous magnetization in YIG. 
\par

%%%DIscussion
As an origin of the frequency shifts in the YIG cantilever, let us first consider magnetostriction effects under external magnetic fields. In the YIG cantilever, the magnetostriction effect might affect the vibrational properties. However, the megnetostriction coefficient for YIG is as small as $10^{-4}$ \% \cite{clark, smith}, and thus the possible frequency shifts due to magnetostriction effects are expected to be much smaller than the observed shifts ($10^{0}$ \%). Therefore, the magnetostriction in the YIG cantilever is not likely to explain the large frequency shifts observed in Figs. \ref{fig3}(a) and (b).
\par

%%%  Figure 4  %%%
\begin{figure}[t]
\begin{center}
\includegraphics[width=8.5cm]{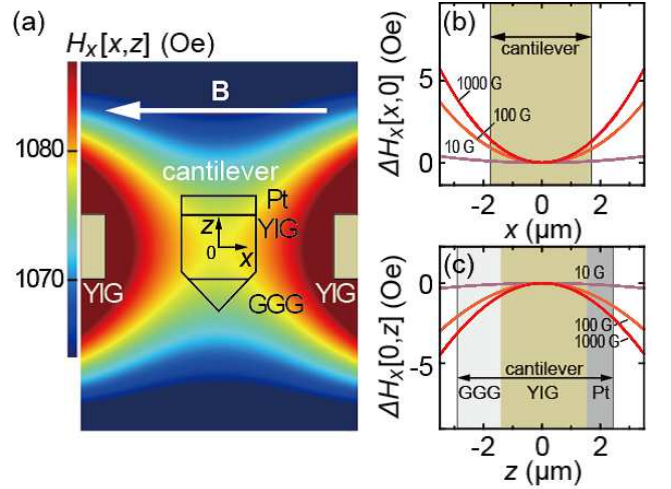}
\caption{(a) Contour plot of stray-magnetic-field profile simulated around the YIG cantilever. The $x$-component of the stray field at the point [$x$, $z$], ($H_{x}$[$x$, $z$]) is mapped. The simulation was performed in the cross section around the head part of the YIG cantilever under $B = 1000$ G, where $B$ is a unidirectional external magnetic field applied along the $x$-axis. (b),(c) The spatial change in the stray field $\Delta H_{x} \equiv H_{x}[x, z] - H_{x}[0, 0]$ (b) in the $x$ direction at $z$ = 0 ($\Delta H_{x}[x, 0] = H_{x}[x, 0] - H_{x}[0, 0]$) and (c) in the $z$ direction at $x = 0$ ($\Delta H_{x}[0, z] = H_{x}[0, z] - H_{x}[0, 0]$). The point at [0, 0] is set at the center of the YIG cantilever (see (a)). $H_{x}$[0, 0] are 7.70 Oe, 101 Oe, and 1030 Oe under $B$ = 10 G, 100 G, and 1000 G, respectively.}
\label{fig4}
\end{center}
\end{figure}

Since the fabricated YIG cantilever is surrounded by the YIG film, spacial gradients of the stray fields around the cantilever should affect the vibrational properties through the magnetic force gradients. When the magnetization of the YIG cantilever is uniformly aligned in the magnetic-field direction (defined as the $x$ direction), the magnetic force gradient for the $i$ direction ($i=x, y, z$)\cite{mfm} is given by
\begin{equation}
\frac{\partial F_{Mag}}{\partial i} = \frac{\partial^{2} ( \vec{M} \cdot \vec{H})}{\partial^{2} i} = M_{x} \frac{\partial^{2} H_{x}}{\partial^{2} i} \equiv k_{\rm eff} , 
\label{1stF_mag}
\end{equation}
where $\vec{M}$ is the magnetization and $\vec{H}$ is the magnetic field. Owing to this change in the effective spring constant $k_{\rm eff}$ of the cantilever by the magnetic force gradient, the frequency shift \cite{dfrequency} is expected to be observed, {\it i.e.}
\begin{equation}
\Delta f = \frac{1}{2 \pi} \Bigl( \sqrt{ \frac{k+k_{\rm eff}}{m_{\rm eff}}  } - \sqrt{ \frac{k}{m_{\rm eff}}  } \Bigr).  
\label{Df}
\end{equation}
Here, $k$ is the spring constant in zero magnetic field and $m_{\rm eff}$ is the effective mass of the cantilever. As shown in eq. (\ref{1stF_mag}) and eq. (\ref{Df}), the frequency shift due to the magnetic force gradient is caused by the changes in $M_{x}$ and $\frac{\partial^{2} H_{x}}{\partial i^{2}}$ with magnetic fields.
\par

We performed a numerical simulation for the magnetic force gradient using COMSOL Multiphysics software. Figure \ref{fig4}(a) shows a contour plot of the simulated magnetic-field profile in the cross section around the tip of the YIG cantilever. Because of the magnetization in the YIG film surrounding the cantilever, the stray field around the cantilever has spatial gradients even under the unidirectional magnetic field. Besides, with increasing magnetic field strength, the stray field along the $x$ direction ($H_{x}$) increases owing to the strong magnetization of the surrounding YIG film. Here, the $z$ axis is defined as the direction normal to the cantilever, and the magnetic field is applied along the $x$ axis. The horizontal and vertical vibrations depend on $M_{x}\frac{\partial^{2} H_{x}}{\partial x^{2}}$ and $M_{x}\frac{\partial^{2} H_{x}}{\partial z^{2}}$, respectively. In the horizontal ($x$) direction, the sign of the stray-field curvature $\frac{\partial^{2} H_{x}}{\partial x^{2}}$ is positive, and its magnitude increases with increasing external magnetic fields, as shown in Fig. \ref{fig4}(b). Thus, according to eq. (\ref{1stF_mag}) and eq. (\ref{Df}), the positive $k_{eff}$ results in the valley-like magnetic-field dependence of the frequency shift, as shown in Fig. \ref{fig3}(e). In contrast, in the vertical ($z$) direction, the magnetic field $H_{x}$ is strongest at $[x, z] = [0, 0]$. In this case, as shown in Fig. \ref{fig4}(c), $\frac{\partial^{2} H_{x}}{\partial z^{2}}$ is negative and decrease with increasing magnetic fields, which turns out to give rise to the peak-like magnetic-field dependence of the frequency shift shown in Fig. \ref{fig3}(e). The full calculation of the magnetic field dependences of $M_{x}$ and $\frac{\partial^{2} H_{x}}{\partial i^{2}}$ $(i=x, z)$ quantitatively explains the magnitudes and the signs of the frequency shifts for the horizontal and vertical modes, as indicated by the solid and dashed curves, respectively, in Fig. \ref{fig3}(e). Hence, the large frequency shifts observed in the YIG cantilever induced by the magnetic fields can be explained by magnetic force gradients produced by the surrounding YIG film.
\par

%%%%%%%%%%%%%%%%%%%%%%%%%%%%%%%%%%%%%%%%%%%%
%%%%%%%%%%%%%%%%%%%%%%%%%%%%%%%%%%%%%%%%%%%%
%%%%%%%%%%%%%%%%%%%%%%%%%%%%%%%%%%%%%%%%%%%%
%\section{Conclusion}
In summary, we have reported on the fabrication and the magnetic control of YIG cantilevers. Under the external magnetic fields, the frequencies of the two resonance modes of the cantilever are shifted clearly; the shifts at $1000$ G reach a few percent. The efficient magnetic control of the resonance frequency is well explained by magnetic force gradients from the surrounding YIG film. Since YIG has been typically used for study of various spin current phenomena, the YIG cantilever would be useful for the mechanical detection of spin currents.
\par
%%%%%%%%%%%%%%%%%%%%%%%%%%%%%%%%%%%%%%%%%%%
%%%%%%%%%%%%%%%%%%%%%%%%%%%%%%%%%%%%%%%%%%%
%%%%%%%%%%%%%%%%%%%%%%%%%%%%%%%%%%%%%%%%%%%
%\section{Acnowledgements}
We thank S. Maekawa, M. Ono, M. Matsuo, Y. Oikawa, and T. Hioki for fruitful discussions. This work was supported by ERATO, Spin Quantum Rectification Project.


\begin{thebibliography}{12}


%%%  Quantum spin liquid and geometrical frustration   %%%
\bibitem{spin-mechanics}J. E. Losby and M. R. Freeman, arXiv:1601.00674 (2016).
\bibitem{spintronics}S. Maekawa, S. O. Valenzuela, E. Saitoh, T. Kimura ,{\it Spin Current}, Oxford University Press, (2012).
\bibitem{EdH}A. Einstein and W. J. de Haas, Royal Neth. Acad. Arts Sciences (KNAW) \textbf{18}, 696 (1915).
\bibitem{Barnett}S. J. Barnett, Phys. Rev. {\bf6}, 239 (1915).
\bibitem{wallis}T. M. Wallis, J. Moreland, and P. Kabos, Appl. Phys. Lett. \textbf{89}, 122502 (2006).
\bibitem{mohanty}P. Mohanty, G. Zolfagharkhani, S. Kettemann, and P. Fulde, Phys. Rev. B {\bf 70}, 195301 (2004).
\bibitem{zolf}G. Zolfagharkhani, A. Gaidarzhy, P. Degiovanni, S. Kettemann, P. Fulde, and P. Mohanty, Nature Nanotech. {\bf 3}, 720-723 (2008).
\bibitem{boales}J. A. Boales, C. T. Boone, and P. Mohanty, Phys. Rev. B {\bf 93}, 161414(R) (2016).
\bibitem{Losby}J. E. Losby, F. Fani Sani, D. T. Grandmont, Z. Diao, M. Belov, J. A. J. Burgess, S. R. Compton, W. K. Hiebert, D. Vick, K. Mohammad, E. Salimi, G. E. Bridges, D. J. Thomson, M. R. Freeman, Science \textbf{350}, 798 (2015).
\bibitem{Wu}M. Wu, N. L.-Y. Wu, T. Firdous, F. F. Sani,	J. E. Losby, M. R. Freeman, and P. E. Barclay, Nature Nanotech.  (2016). 
\bibitem{stowe}T. D. Stowe, K. Yasumura, T. W. Kenny, D. Botkin, K. Wago, and D. Rugar, Appl. Phys. Lett. \textbf{71},  288-290 (1997).
\bibitem{chaste}J. Chaste, A. Eichler, J. Moser, G. Ceballos, R. Rurali, and A. Bachtold, Nature Nanotech. \textbf{7}, 301-304 (2012).
\bibitem{zhang}Z. Zhang, P. C. Hammel, and P. E. Wigen, Appl. Phys. Lett. \textbf{68}, 2005 (1996).
\bibitem{wago}K. Wago, D. Botkin, C. S. Yannoni, and D. Rugar, Appl. Phys. Lett. \textbf{72}, 2757 (1998).
\bibitem{midzor}M. M. Midzor, P. E. Wigen, D. Pelekhov, W. Chen, P. C. Hammel, and M. L. Roukes, J. Appl. Phys. \textbf{87}, 6493 (2000).
\bibitem{charbois}V. Charbois, V. V. Naletov, J. B. Youssef, and O. Klein, Appl. Phys. Lett. {\bf 80}, 4795 (2002).
\bibitem{naletov}V. V. Naletov, V. Charbois, O. Klein, and C. Fermon, Appl. Phys. Lett. \textbf{83}, 3132 (2003).
\bibitem{klein}O. Klein, V. Charbois, V. V. Naletov, and C. Fermon, Phys. Rev. B {\bf 67}, 220407 (2003).
\bibitem{comsol}https://www.comsol.com/
\bibitem{elastic1}D. F. Gibbons and V. G. Chirba, Phys. Rev. \textbf{110}, 770-771 (1958).
\bibitem{elastic2}H. M. Chou and E. D. Case, Materials Science and Engineering, \textbf{100}, 7-14 (1988).
\bibitem{mforce}J. A. Sidles, J. L. Garbini, K. J. Bruland, D. Rugar, O. Zuger, S. Hoen, and C. S. Yannoni, Rev. Mod. Phys. \textbf{67}, 249-268 (1995).
\bibitem{mforce2}U. Durig, O. Zuger, and A. Stalder, J. Appl. Phys. \textbf{72}, 1778 (1992).
\bibitem{clark}A. E. Clark, B. DeSavage, W. Coleman, E. R. Callen and H. B. Callen, J. Appl. Phys. \textbf{34}, 1296-1297 (1963).
\bibitem{smith}A. B. Smith and R. V. Jones, J. Appl. Phys. \textbf{34}, 1283-1284 (1963).
\bibitem{mfm}K. Babcock and V. Elings, IEEE Trans. Magn., \textbf{30}, 4503-4505 (1994).
\bibitem{dfrequency}A. M\'endez-Vilas and J. D\'iaz, {\it Modern Research and Educational Topics in Microscopy}, FORMATEX 805-811 (2007).





\end{thebibliography}
\end{document}